\documentclass{article}
\usepackage{graphicx}
\begin{document}

\newcommand{\be}{\begin{equation}}
\newcommand{\ee}{\end{equation}}

\newcommand{\bn}{\begin{eqnarray}}
\newcommand{\en}{\end{eqnarray}}

\newcommand{\gh}{\mbox{$SU(3)_H$}}

\title{ Radiative generation of light fermion masses in a
$SU(3)_{H}$ horizontal symmetry model}

\author{Albino Hern\'andez Galeana \\
\small Departamento de F\'{\i}sica,\\
\small  Escuela Superior de F\'{\i}sica y Matem\'aticas, I.P.N.,\\
\small  U. P. "Adolfo L\'opez Mateos". C. P. 07738, M\'exico,
D.F., M\'exico}

\maketitle

\begin{abstract}

In a model with a gauge group $SU(3)_H\otimes G_{SM}$, where
$SU(3)_H$ is a horizontal symmetry and $G_{SM}=SU(3)_C\otimes
SU(2)_L\otimes U(1)_Y$ is the standard model, we propose a
radiative mechanism of mass generation mediated by the $SU(3)_H$
gauge bosons for the light fermions, meanwhile the masses of the
heaviest family are generated by the implementation of see-saw
mechanisms with the introduction of vectorial fermions.

\end{abstract}

Keywords: Horizontal symmetry; fermion masses

\vspace{5mm}

\noindent{\footnotesize  PACS: 12.15.Ff; 12.10.-g}

\section{ Introduction }

 The known quark and lepton masses are generated after the spontaneous breaking
 of the electroweak symmetry to $U(1)_Q$  of Quantum Electrodynamics. However,
 the responsible mechanism of this symmetry breaking as well as the generation
 mechanism of the fermion masses, including their origin and hierarchy, remain
 unknown and they are two of the current puzzles of the elementary particle
 physics. In the literature there are many proposals
trying to explain the mass hierarchy, the fermion mixing and their
possible relation to new physics [1].

A possible answer to why the masses of the light fermions are so
small compared with electroweak scale is that they arise through
radiative corrections [ 2 ], while the mass of the top quark and
probably those of the bottom quark and of the tau lepton are
generated at tree level. This may be understood as a consequence
of the breaking of a symmetry among families ( a horizontal
symmetry ). This symmetry may be discrete [3 ], or continuous, [ 4
]. The radiative generation of the light fermions may be mediated
by scalar particles as it is proposed, for instance, in references
[ 2 ] and [ 5 ] or also through vectorial bosons as it happens in
"Dynamical Symmetry Breaking" (DSB) theories like " Extended
Technicolor ", [ 6 ].

In this work we propose a mechanism in which the light fermions
get mass through loop diagrams mediated by the massive vectorial
bosons of a   horizontal symmetry that is spontaneously broken,
whereas the masses of the top and bottom quarks as well as the tau
lepton are generated by the implementation of see - saw mechanisms
with the introduction of new fermions of vectorial type.

\section{ The model }

We define the gauge group symmetry $G\equiv SU(3)_H \otimes
SU(3)_C \otimes SU(2)_L \otimes U(1)_Y$ , where $SU(3)_H$ is a
horizontal symmetry among families and $G_{SM}\equiv SU(3)_C
\otimes SU(2)_L \otimes U(1)_Y$ is the gauge group of the
"Standard Model" of elementary particles. The content of fermions
assume the ordinary quarks and leptons including the sector of
right-handed neutrinos, assigned under the G group in the form:

\be
\Psi_q = ( 3 , 3 , 2 , \frac{1}{3} )_L
\ee

\be
\Psi_u = ( 3 , 3, 1 , \frac{4}{3} )_R \ee

\be
\Psi_d = (3, 3 , 1 , -\frac{2}{3} )_R \ee

\be
\Psi_l = ( 3 , 1 , 2 , -1 )_L \ee

\be
\Psi_\nu = ( 3 , 1 , 1 , 0 )_R
\ee

\be
\Psi_e = (3 , 1 , 1,-2)_R
\ee

\noindent
 where the last entry correspond to the  hypercharge
$(Y)$ and the electric charge is defined by $Q = T_{3L} +
\frac{1}{2} Y$ . The model also includes the introduction of the
following vectorial fermions:

\begin{equation}
U_L, U_R = ( 1 , 3 , 1 , \frac{4}{3} )
\end{equation}

\begin{equation}
D_L, D_R = ( 1 , 3 , 1 ,- \frac{2}{3} )
\end{equation}

\begin{equation}
N_L, N_R = ( 1 , 1 , 1 , 0 )
\end{equation}

\begin{equation}
E_L, E_R = ( 1 , 1 , 1 , -2 )
\end{equation}

The above fermionic content  and its assignment under the G group
make the model  anomaly free. At this point we consider convenient
to explain briefly about the reasons for the introduction of the
right-handed neutrinos and the vectorial fermions. After the
definition of the gauge group G and the assignment of the ordinary
fermions in the canonical form under the standard model group, in
the most simply non trivial way under the horizontal symmetry, the
introduction of the right-handed neutrinos becomes a necessity to
cancel anomalies, while the introduction of the vectorial fermions
has its origin in the procedure to give masses at tree level only
to the heaviest family of known fermions. We consider that these
vectorial fermions play a crucial role to implement our proposal
hierarchical mass generation mechanism.

\section{ Symmetry breaking}

 The "Spontaneous Symmetry Breaking" (SSB) is proposed to be achieved in the form:

\be
G \stackrel{\Lambda_1}{\longrightarrow} G_{SM}
\stackrel{\Lambda_2}{\longrightarrow} SU(3)_C \otimes U(1)_Q
\ee

\noindent
 in order to the model had possibility to be consistent
with the known low energy physics,  where $\Lambda_1$   and
$\Lambda_2$ are the scales of SSB.

With the intention to implement the mass generation mechanism for
the ordinary quarks and leptons and simultaneously to contribute
to the SSB of $\gh$ at the first stage, we introduce the scalar
field:

\be
\Phi^{\prime \prime} = ( 3 , 1 , 1 , 0 ),
\ee

\noindent
 with "Vacuum Expectation Value" (VEV),

\be < \Phi^{\prime \prime} >^T = ( 0 , 0 , v^{\prime \prime} ) \ee

\noindent
 where $T$ means transpose. This scalar $\Phi^{\prime
\prime}$ with the above VEV generates the breaking of $\gh$  to
$SU(2)_H$ . Therefore, to complete the breaking of the horizontal
symmetry and to produce the appropriate mixing among their neutral
gauge bosons is necessary to introduce more scalar fields. The
details to achieve this goal are not treated in this paper.
However, we can mention here that the additional Higgs scalars
should not be in the fundamental representation of $\gh$  in order
they do not spoil the desired hierarchical fermion mass generation
mechanism.

\vspace{3mm}
 To achieve the spontaneous breaking of the electroweak symmetry to $U(1)_Q$,
  we introduce the scalars:
\be
\Phi = ( 3 , 1 , 2 , -1 )
\ee

\be
\Phi^{\prime} = ( 3 , 1 , 2 , +1 )
\ee

\noindent with the VEV$^{\prime}$s

\be
< \Phi >^T = ( < \Phi_1 > , < \Phi_2 > , < \Phi_3 > )\;\; ; \;\;\;
< \Phi_i
> = \frac{1}{\sqrt[]{2}} \left( \begin{array}{c} v_i
\\ 0  \end{array} \right)
\ee

\be
< \Phi^{\prime} >^T = ( < \Phi^{\prime}_1 > , < \Phi^{\prime}_2 >
, < \Phi^{\prime}_3 > ) \;\; ; \;\;\; < \Phi^{\prime}_i > =
\frac{1}{\sqrt[]{2}} \left( \begin{array}{c} 0 \\ V_i  \end{array}
\right)
\ee

\noindent
where $i = 1,2,3$. The contributions of $< \Phi >$ and
$< \Phi^{\prime} >$ to the mass of the $W$ charged gauge boson
are:

\noindent
contribution of $< \Phi >$ :
\be
\frac{1}{2} g ( v^{2}_1 + v^{2}_2 + v^{2}_3 )^{\frac{1}{2}}
\ee

\noindent
contribution of $< \Phi^{\prime} >$ :
\be
\frac{1}{2} g ( V^{2}_1 + V^{2}_2 + V^{2}_3 )^{\frac{1}{2}}
\ee

\noindent where g is the $SU(2)_L$   coupling constant. If now we
take into account that the horizontal symmetry has already been
completely broken at the first stage and if $\Lambda_1 >>
\Lambda_2$, then there are not arguments of symmetry to align the
VEV$^{\prime}$s of $\Phi$ and $\Phi^{\prime}$ under $\gh$ , that
is, we can assume for simplicity that

\be
v_1 = v_2 = v_3 = V_1 = V_2 = V_3 \equiv \frac{v}{ \sqrt[]{6}}
\ee

\noindent where $v \simeq 246 \; GeV$ and then $M_W = \frac{1}{2}
gv$. Note that $\Phi$ and $\Phi^{\prime}$ transform as triplets
under $\gh$. This fact has as consequences that they contribute
slightly to the $\gh$ gauge boson masses and produce mixing
between these bosons with the neutral boson Z of the standard
model. The consequences of this phenomenon will not be discussed
in this report.

\section{ Fermion masses}

Now we describe briefly how to obtain the mass terms of the
ordinary fermions. The analysis is presented  for the up-quark
sector, with a completely analogous procedure for the other
sectors.

With the fields of particles introduced in the model, we may write
the fo\-llowing gauge invariant  Yukawa  couplings:

\be h_u \bar{\Psi}_q \Phi U_R \;\;+\;\; h^{\prime \prime}
\bar{\Psi}_u \Phi^{\prime \prime} U_L \;\;+\;\; M \bar{U}_L U_R
\;\;+ h.c \ee

\noindent where $h_u$ and $h^{\prime \prime}$ are Yukawa coupling
constants.

 When $\Phi$    and $\Phi^{\prime \prime}$  acquire VEV$^{\prime}$s we
 obtain the mass terms

\be
\bar{\Psi}^{o}_L M^{o}_u \Psi^{o}_R  \;\;+ h.c \ee

\noindent
where:

\be
{\Psi^{o}_L}^T = ( u^{o}_L , c^{o}_L , t^{o}_L , U_L )
\ee

\be
{\Psi^{o}_R}^T = ( u^{o}_R , c^{o}_R , t^{o}_R , U_R )
\ee

\noindent
and

\be M^{o}_u = \left( \begin{array}{cccc} 0 & 0 & 0 & hv\\ 0 & 0 &
0 & hv\\ 0 & 0 & 0 & hv\\ 0 & 0 & h^{\prime \prime} v^{\prime
\prime} & M
\end{array} \right)
\ee

\noindent with $h = \frac{1}{ \sqrt[]{12}} h_u$. When this mass
matrix $M^{o}_u$ is diagonalized by a biunitary transformation we
find to this order two associated zero eigenvalues  to the masses
of the light up quarks, while the two eigenvalues different from
zero are associated with the masses of the top quark and the
vectorial fermion U.

Explicitly, if we define

\be
\Psi^{o}_L = V^{o}_L \Psi_L
\ee

\be
\Psi^{o}_R = V^{o}_R \Psi_R
\ee

\be
\lambda_{\pm } = \frac{1}{2} \left( B \pm \sqrt[]{B^2 -4D} \right)
\ee

\noindent
where

\be
B = 3 a^2 + b^2 + c^2
\ee

\be
D = 3 a^2 b^2
\ee

\noindent
with

\be a \equiv hv \;\;,\;\; b \equiv h^{\prime \prime} v^{\prime
\prime} \;\;,\;\;c \equiv M \ee

\vspace{3mm}
\noindent
the orthogonal matrices $V^{o}_L$   and
$V^{o}_R$ become

\be
V^{o}_L = \left( \begin{array}{cccc} \frac{1}{\sqrt[]{2}} &
\frac{1}{\sqrt[]{6}} & \frac{1}{\sqrt[]{3}} \cos{\alpha} &
\frac{1}{\sqrt[]{3}} \sin{\alpha}\\ -\frac{1}{\sqrt[]{2}} &
\frac{1}{\sqrt[]{6}} & \frac{1}{\sqrt[]{3}} \cos{\alpha} &
\frac{1}{\sqrt[]{3}} \sin{\alpha}\\ 0 & -\frac{2}{\sqrt[]{6}} &
\frac{1}{\sqrt[]{3}} \cos{\alpha} & \frac{1}{\sqrt[]{3}}
\sin{\alpha}\\ 0 & 0 & -\sin{\alpha} & \cos{\alpha}  \end{array}
\right)
\ee

\noindent
where

\be
\cos{\alpha} = \frac{(\lambda_+ - 3 a^2)}{\sqrt[]{ 3 a^2 c^2 +
(\lambda_+ - 3 a^2)^2}}
\ee

\be
\sin{\alpha} = \frac{\sqrt[]{3} a c}{\sqrt[]{ 3 a^2 c^2 +
(\lambda_+ - 3 a^2)^2}}
\ee

\noindent
and

\be
V^{o}_R = \left( \begin{array}{cccc} 1 & 0 & 0 & 0\\ 0 & 1 & 0 &
0\\ 0 & 0 & \cos{\beta} & \sin{\beta}\\ 0 & 0 & - \sin{\beta} &
\cos{\beta}  \end{array}  \right)
\ee

\noindent
with

\be
\cos{\beta} = \frac{(\lambda_+ - b^2)}{\sqrt[]{b^2 c^2 +
(\lambda_+ - b^2)^2}}
\ee

\be
\sin{\beta} = \frac{b c}{\sqrt[]{b^2 c^2 + (\lambda_+ - b^2)^2}}
\ee

\noindent
From these matrices $V^{o}_L$    and $V^{o}_R$   we
obtain

\be
{V^{o}_L}^T M^{o}_u V^{o}_R = \left( \begin{array}{cccc} 0 & 0 & 0
& 0\\ 0 & 0 & 0 & 0\\ 0 & 0 & - \; \sqrt[]{\lambda_{-}} & 0\\ 0 &
0 & 0 & \sqrt[]{\lambda_+} \end{array}  \right) \ee

\noindent
and the top-quark mass is given approximately by

\be m_{t} \simeq \sqrt[]{3} h h^{\prime \prime} \frac{v v^{\prime
\prime}}{M_{U}} \ee

\noindent
$M_{U}$ being the mass of the vectorial quark U.

\vspace{5mm} Subsequently the masses of the light up quarks arise
through one and two loop diagrams. After the breakdown of the
electroweak symmetry, we can construct the generic  one loop mass
diagram of Figure 1. The vertices in this diagram come from the
gauge interaction Lagrangian

\be i L_{int} = \frac{g_{H}}{\sqrt[]{2}} \left(
\overline{u^{o}}_{kL} \gamma_{\mu} u^{o}_{jL} Y^{\mu}_{kj} +
\overline{u^{o}}_{iR} \gamma_{\mu} t^{o}_R {Y^{+}_{it}}^{\mu}
\right) + L \leftrightarrow R  + h.c \ee

\noindent $g_H \;(Y)$ being the $\gh$  coupling constant ( gauge
bosons ), $i, j, k = 1, 2, 3$ denote family indices, the crosses
in the internal fermion line mean the mixing, and the mass $M$,
generated by the couplings of eq. (21) after $\Phi$  and
$\Phi^{\prime \prime}$ take VEV$^{\prime}$s and the black circle
in the boson line means the tree level mixing mass term

\be
M^2_{ij} Y^{+}_{ik} Y_{tj}
\ee

\noindent
generated in the  symmetry breaking process.

\section{ Discussion}

At present we are performing a more complete study of the model
introduced in this report, including some phenomenology and
confrontation with the experimental data. For the last, an
important aspect to study is the calculation of some processes of
"Flavor Changing Neutral Currents" (FCNC) that  can be induced in
the model as a consequence of the particles introduced, for
instance the gauge bosons of the $\gh$  horizontal symmetry.  Some
examples of these FCNC processes are the radiative "lepton flavor
violation" (LFV)  decays: $\mu \rightarrow e \gamma$ , $\tau
\rightarrow \mu \gamma $ and $\tau \rightarrow e \gamma $ , which
are induced by the radiative mass generation mechanism when we
attach a photon to the internal charged leptonic lines in the one
loop mass diagrams, and $\mu \rightarrow e e e $, $\tau
\rightarrow \mu \mu \mu $, $\tau \rightarrow \mu \mu e $ , $\tau
\rightarrow \mu e e $ and $\tau \rightarrow eee $, coming from
tree level diagrams like the one in Figure 2. The calculation
involved, including the muon anomalous magnetic moment induced as
well by the radiative muon mass generation mechanism [ 7 ],  is in
progress and will be reported elsewhere.

\vspace{5mm}
Let us remark here the interesting feature of the
model concerning the introduction of the sector of right-handed
neutrinos as a requirement to cancel anomalies. This fact allow us
to generate radiatively Dirac type neutrino masses and then to
study the current and recent interesting phenomena concerning
neutrino physics, such as neutrino oscillations, neutrino masses,
mixing and so on [ 8 ].

\vspace{3mm}
Finally, notice that eq.(39) can be used to write the
mass of the vectorial-quark U as

\be M_U \simeq \sqrt[]{3} h h^{\prime \prime} \frac{v}{m_t}
v^{\prime \prime} \ee

\noindent If in this last equation we consider the values  $m_t
\simeq 175 \; GeV$ and $v \simeq 246 \; GeV$, then by assuming
$\sqrt[]{3} h h^{\prime \prime} \simeq \frac{m_t}{v} \simeq .711$
we could expect $M_U$ to be of the same order of $v^{\prime
\prime}$, that is, of the order of the scale of SSB of $\gh$ that
could work in the $TeV$ region. This order of magnitude may also
be expected for the other vectorial fermion masses if one supposes
that the corresponding products of Yukawa coupling constants
behave like $\sqrt[]{3} h_d h^{\prime \prime}_d \simeq
\frac{m_b}{v}$ and $\sqrt[]{3} h_e h^{\prime \prime}_e \simeq
\frac{m_{\tau}}{v}$. However, the actual values of the Yukawa
coupling constants, the VEV $v^{\prime \prime}$ and the masses of
the vectorial fermions involved should be determined by demanding
consistence between the fermion masses and mixing angles with the
adequate suppression of FCNC ( if this suppression is realized).

 \vspace{5mm}

Acknowledgments

\vspace{5mm} This work was partially supported by the "Instituto
Polit\'ecnico Nacional", (Grants from EDI and COFAA) and SNI in
Mexico.

\section{ References}

{\bf [1] } H. Fritzsch and Z. Xing, Prog. Part. Nucl. Phys. 45
(2000) 1,\\ hep-ph/9912358. E. Ma, hep-ph/0401044.

A. Hern\'andez-Galeana, W. A. Ponce and A. Zepeda, Zeitschrift Fur
Physik\\  C, Vol. 55 (3), (1992) 423-434.

J. Kubo, A. Mondragon, M. Mondragon, E. Rodriguez-Jauregui, Prog.
Theor.\\ Phys. 109 (2003) 795-807. Mondragon, E.
Rodriguez-Jauregui, Rev. Mex. Fis. 46 (2000) 5-22,
       Phys. Rev. D61 (2000) 113002, Phys. Rev. D59 (1999) 093009,  hep-ph/9807214.

H. Abele, E. Barberio, D. Dubbers, F. Glueck, J. C. Hardy, W. J.
Marciano, A. Serebrov, N. Severijns,  hep-ph/0312150.

\vspace{3mm}
 {\bf [2] }X.G. He,
R. R: Volkas, and D. D. Wu, Phys. Rev. D 41 (1990) 1630; Ernest
Ma, Phys. Rev. Lett. 64 (1990) 2866.

\vspace{3mm} {\bf [3] } Sandip Pakvasa and Hirotaka Sugawara,
Phys. Lett. B73 (1978) 61; Y. Yamanaka, H. Sugawara, and S.
Pakvasa, Phys. Rev. D 25 (1982) 1895; K. S. Babu and X.G. He,
ibid. 36 (1987) 3484; Ernest Ma, Phys. Rev. Lett. B 62 (1989) 61.

\vspace{3mm} {\bf [4] } A. Davidson, M. Koca, and K. C. Wali,
Phys. Rev. Lett. B 43 (1979) 92, Phys. Rev. D 20 (1979) 1195; C.
D. Froggatt and H. B. Nielsen, Nucl. Phys. B 147 (1979) 277; A.
Sirlin, Phys. Rev. D 22 (1980) 971; A. Davidson and K. C. Wali,
ibid. 21 (1980) 787.

\vspace{3mm} {\bf [5] } E. Garcia, A. Hernandez-Galeana, D.
Jaramillo, W. A. Ponce and A. Zepeda, Revista Mexicana de Fisica
Vol. 48(1), (2002) 32, hep-ph/0006093; E. Garcia, A.
Hernandez-Galeana, A. Vargas and A. Zepeda, hep-ph/0203249.

\vspace{3mm} {\bf [6] } For reviews, see E. H. Simmons, in Proc.
of the APS/DPF/DPB summer study on the future of Particle Physics
(Snowmass 2001) ed. N. Graf, eConf C010630, P110 (2001) C. T. Hill
and E. H. Simmons, Phys. Rept. 381, 235 (2003). T. Appelquist, M.
Piai  and  R. Shrock, hep-ph/0401114, hep-ph/0308061. T.
Appelquist, R. Shrock, Phys. Rev. Lett. 90 (2003) 201801.

See also:

A. Hernandez-Galeana and A. Zepeda, Revista Mexicana de Fisica,
Vol. 45(3),(1999)239.

A. Hernandez-Galeana and A. Zepeda, edited by A. Ali and P.
Hoodbhoy, (World Scientific, 1991), in memory of Prof. M.A.B.
B\'eg., pag.102.

A. Hernandez-Galeana, Ph. D. Thesis, Centro de Investigaci\'on y
de Estudios Avanzados del IPN, (Mexico, 1989).

A. Hernandez-Galeana and A. Zepeda, Zeitschrift fur Physik C, Vol.
40,(1988)125.

M.A.B. B\'eg, in {\it Mexican School of Particles and Fields},
edited by J.L. Lucio, A. Zepeda and M. Moreno (AIP, New York,
1986). Conference Proceedings No. 143, pag. 1.

A. Zepeda, Phys. Lett. 132 B (1983) 407; (1987) 623E.

\vspace{3mm} {\bf [7] } W. J. Marciano, in {\it Particle Theory
and Phenomenology }, edited by K. Lassila {\it et al}. (World
Scientific, Singapore, 1996), p 22;

A. Czarnecki and W. J. Marciano, Phys. Rev. D64 (2001) 013014,
hep-ph/0102122; A. Czarnecki and W. J. marciano, hep-ph/0010194.

\vspace{3mm} {\bf [8] } J. A. Aguilar-Saavedra, G. C. Branco and
F. R. Joaquin, hep-ph/0310305.

S. Palomares-Ruiz and J. Bernabeu, hep-ph/0312038, hep-ph/0311354.

E. Ma, hep-ph/0405152, hep-ph/0401025, hep-ph/0312192,
hep-ph/0311215, hep-ph/0308282, hep-ph/0404199.

C. P. Burgess, N. S. Dzhalilov, M. Maltoni, T. I. Rashba, V.B.
Semikoz, M. Tortola,

J. W. F. Valle, hep-ph/0312345.

M. Hirsch, J. C. Romao, S. Skadhauge, J. W. F. Valle, A. Villanova
del Moral, hep-ph/0312265.

O. G. Miranda, T. I. Rashba, A. I. Rez, J. W. F. Valle,
hep-ph/0311014.

J. W. F. Valle, hep-ph/0310125, hep-ph/0307192.

A.Yu. Smirnov, hep-ph/0402264, hep-ph/0311259.

O. L. G. Peres, A. Yu. Smirnov, hep-ph/0309312.

P. C. de Holanda, A. Yu. Smirnov, hep-ph/0309299.

J. Bernabeu, S. Palomares-Ruiz, S. T. Petcov, Nucl. Phys. B669
(2003) 255-276.

H. S. Goh, R. N. Mohapatra, S. Nasri, Siew-Phang Ng,
hep-ph/0311330.

H. S. Goh, R. N. Mohapatra, Siew-Phang Ng, Phys. Rev. D68 (2003)
115008.

R. N. Mohapatra, hep-ph/0402035, hep-ph/0306016, hep-ph/0211252.

A. Dutta, R. N. Mohapatra, Phys. Rev. D68 (2003) 056006.

H. S. Goh, R. N. Mohapatra, Siew-Phang Ng, Phys. Lett. B570 (2003)
215-221.

R. N. Mohapatra, A. Perez-Lorenzana, Phys. Rev. D67 (2003) 075015.

K. Fujikawa, R. Shrock, hep-ph/0303188.

T. Appelquist, R. Shrock, Phys. Lett. B548 (2002) 204-214.

W. J. Marciano, hep-ph/0108181.

B. Bajc, G. Senjanovic, F. Vissani, hep-ph/0402140, Phys. Rev.
Lett. 90 (2003) 051802, hep-ph/0210207.

G.L. Fogli, E. Lisi, A. Marrone, A. Melchiorri, A. Palazzo, P.
Serra, J. Silk, hep-ph/0408045.

K. Fujikawa, hep-ph/0407331.

A. Masiero, S.K. Vempati, O. Vives, hep-ph/0407325.

J.C. Pati, hep-ph/0407220.

R.N. Mohapatra, S. Nasri, hep-ph/0407194.

A. Kopylov, hep-ph/0407184.

H. Athar, C.S. Kim, hep-ph/0407182.

A. Strumia, hep-ph/0407132.

S.M. Bilenky, hep-ph/0407125.

S.L. Glashow, hep-ph/0407087.

S. Gardner, V. Bernard, U.G. Mei{it B}ner, hep-ph/0407077.

J.C. Montero, V. Pleitez, M.C. Rodriguez, hep-ph/0406299.

J.N. Bahcall, M.C. Gonzalez-Garcia, C. Pe'~na-Garay,
hep-ph/0406294.

O.G. Miranda, M.A. Tortola, J.W.F. Valle, hep-ph/0406280.

Wan-lei Guo, hep-ph/0406268.

B.C. Chauhan, J. Pulido, hep-ph/0406227.

H. Fritzsch, Zhi-zhong Xing, hep-ph/0406206.

P. Kaus, S. Meshkov, hep-ph/0406195.

K.M. Beshtoev, hep-ph/0406124.

P. Di Bari, hep-ph/0406115.

S.T. Petcov, S. Palomarez-Ruiz, hep-ph/0406106.

S. Palomarez-Ruiz, S.T. Petcov, hep-ph/0406096.

J.P. Archambault, A. Czarnecki, M. Pospelov, hep-ph/0406089.

K.M. Beshtoev, hep-ph/0406084.

M.C. Gonzalez-Garcia, M. Maltoni, hep-ph/0406056, hep-ph/0404085.

S. Raby, hep-ph/0406022.

A. Broncano, M.B. Gavela, E. Jenkins, hep-ph/0406019.

R. Dermisek, hep-ph/0406017.

Riazuddin, hep-ph/0405289.

W. Grimus, L. Lavoura, hep-ph/0405261.

C. Boehm, hep-ph/0405240.

K.S. Babu, hep-ph/0405197.

J. Hisano, hep-ph/0405185.

G. Altarelli, hep-ph/0405182.

S. Pakvasa, hep-ph/0405179.

M. Maltoni, T. Schwetz, M.A. Tortola, J.W.F. Valle,
hep-ph/0405172.

S. Antusch, S.F. King, hep-ph/0405093.

G. Marandella, hep-ph/0405090.

H. Minakata, A.Y. Smirnov, hep-ph/0405088.

G. Altarelli, F. Feruglio, hep-ph/0405048.

S. Godfrey, S. Zhu, hep-ph/0405006.

G. Gelmini, G. Varieschi, T. Weiler, hep-ph/0404272.

S. Hannestad, hep-ph/0404239.

B. Kyae, Q. Shafi, hep-ph/0404168.

F. Buccella, D. Falcone, hep-ph/0404159.

C.A. de S. Pires, hep-ph/0404146.

G.K. Leontaris, J. Rizos, A. Psallidas, hep-ph/0404129.

J.A. Lopez-Perez, N. Rius, hep-ph/0404124.

J.N. Bahcall, C. Pe\~na-Garay, hep-ph/0404061.

D. Kazanas, R.N. Mohapatra, S. Nasri, V.L. Teplitz,
hep-ph/0403291.

P.F. Harrison, W.G. Scott, hep-ph/0403278.

N. Cosme, hep-ph/0403209.

W.J. Marciano, Z. Parsa, hep-ph/0403168.

P.H. Frampton, hep-ph/0403164.

M.M. Guzzo, P.C. de Holanda, O.L.G. Peres, hep-ph/0403134.

A. Friedland, C. Lunardini, C. Pe\~na-Garay, hep-ph/0402266.

M. Barkovich, J.C. D'Olivo, R. Montemayor, hep-ph/0402259.

A. Romanino, hep-ph/0402258.

C. Giunti, hep-ph/0402217.

L. Oberauer, hep-ph/0402162.

G. Altarelli, F. Feruglio, I. Masina, hep-ph/0402155.

R.D. McKeown, P. Vogel, hep-ph/0402025.

P.H. Frampton, S.T. Petcov, W. Rodejohann, hep-ph/0401206.

W. Krolikowski, hep-ph/0401101.

M. Maltoni, hep-ph/0401042.

\newpage

\begin{figure}[top]
\begin{center}
\includegraphics{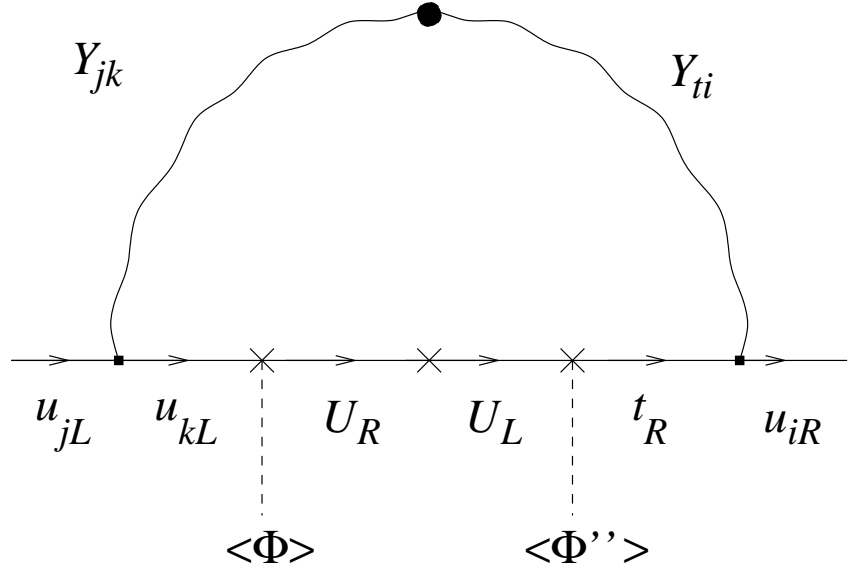}
\end{center}
\caption{\label{fig1} One loop diagram contribution to the mass
term $m_{ij} \bar{u}_{iR} u_{jL}$}
\end{figure}

\vspace{5mm}

\begin{figure}[htp]
\begin{center}
\includegraphics{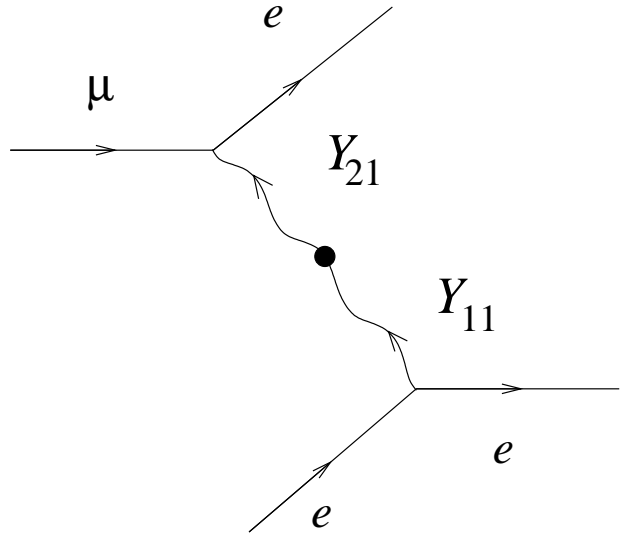}
\end{center}
\caption{\label{fig2} Tree level diagram contribution to the LFV
process $\mu \rightarrow e e e $}
\end{figure}

\end{document}